\documentclass[aps,prd,twocolumn,groupedaddress]{revtex4-1}
\usepackage{graphicx}
\include{amssym}

\begin{document}
\title{$1/N_c$ approximation and universality of vector mesons}

\author{M.\,K. Volkov$^{1}$}
\email[]{volkov@theor.jinr.ru}

\author{A.\,A. Osipov$^{1}$}
\email[]{aaosipov@jinr.ru}

\author{A.\,A. Pivovarov$^{1}$}
\email[]{tex$_$k@mail.ru}

\author{K. Nurlan$^{1,2,3}$}
\email[]{nurlan.qanat@mail.ru}

\affiliation{$^1$Bogoliubov Laboratory of Theoretical Physics, Joint Institute for Nuclear Research, Dubna, 141980, Russia}

\affiliation{$^2$ Institute of Nuclear Physics, 1 Ibragimova, 050032, Almaty, Kazakhstan}

\affiliation{$^3$L.\,N. Gumilyov Eurasian National University, 2 Satpaev, 010000, Nur-Sultan, Kazakhstan}

\begin{abstract}
In this work, we argue that the observed differences in the value of the vector coupling constant extracted from the decays $\rho\to \pi\pi$ ($g_{\rho} = 6.0$), $\rho\to l^+l^-$ ($g_{\rho} = 5.0 $) and $\omega\to l^+ l^-$ ($g_{\rho} = 5.7$), where $l=e,\mu $, are an indication of the important role played by the $1/N_c$ corrections in the description of these processes. We show that an emission of a photon by charged meson loops in the $\rho^0, \omega,\phi\to\gamma$ transitions is a key process that allows to describe above vector meson decays into two leptons with a single value $g_\rho = 6.0$. Our result supports the idea of universality of neutral vector mesons and clarifies the role of accounting of $1/N_c$ corrections to its fulfilment. 
\end{abstract}

\maketitle

\section{Introduction}
It is well-known that the hypothesis of vector-meson dominance (VMD) \cite{Sakurai:69,Feinman:72} possesses one to compute the lepton-pair decay rate of the neutral vector meson $V=\rho^0,\omega,\phi$ in terms of the vector-meson coupling constant $f_V$ appearing in a current-field identity $J_\mu^{em}=\sum_V \frac{m^2_V}{f_V} V_\mu$. This yields 
\begin{equation}
\label{VLL}
\Gamma_{V\to l^+l^-} =\frac{4\pi\alpha^2}{3f_V^2}\left(1+\frac{2m_l^2}{m_V^2}\right)\sqrt{m_V^2-4m_l^2},
\end{equation}  
where $\alpha =e^2/4\pi =1/137$. Current experimental data on $V\to l^+l^-$ decay widths \cite{PDG:20} allow to determine from (\ref{VLL}) the phenomenological values of these constants. For instance, considering the $V\to e^+ e^-$ mode one finds $f_\rho\equiv g_\rho =4.96, \, f_\omega = 17.06, \, f_\phi = -13.44$. On the other hand, the complete $\rho$ dominance of the electromagnetic form factor of $\pi^\pm$ implies $g_\rho =g_{\rho\pi\pi}$ \cite{Sakurai:68}, where the coupling constant of the $\rho\to\pi\pi$ decay is known to be $g_{\rho\pi\pi}=5.96$, as it follows from theoretical decay width of the $\rho$-meson
\begin{equation}
\Gamma_{\rho\to\pi\pi} =\frac{g^2_{\rho\pi\pi}}{48\pi }\, m_\rho \left(1 - \frac{4m^2_\pi}{m^2_\rho} \right)^\frac{3}{2},
\end{equation}
and the phenomenological value $\Gamma_{\rho\to\pi\pi}^{exp} =149.1\pm 0.8$ MeV. We see that describing the strong decay $\rho\to\pi\pi$, one obtains the value $g_\rho \simeq 6.0$. At the same time, when describing electromagnetic decays $\rho\to e^+ e^-$, one obtains $g_\rho\simeq 5.0$, and when describing a related process $\omega\to e^+ e^-$, one finds $g_\rho = f_\omega /3\simeq 5.7$. What is the reason of such differences in extracted values of $g_\rho$? 
 
Our expectation is based on the Sakurai idea of vector meson universality, which suggests to consider the vector mesons as a gauge bosons of a local isospin symmetry \cite{Sakurai:60} with universal coupling $g_\rho$ of these gauge boson to conserved currents. In accord with this idea, kinetic terms  and couplings of the rho-mesons with matter fields have the Yang-Mills form, except a non-zero mass of vector mesons. This idea was extensively studied at the tree-level order in the framework of the "massive Yang-Mills" \cite{Gasiorowicz:69,Schechter:84a,Schechter:84b,Schechter:85} and "hidden-gauge" theories \cite{Bando:88}. There are also reviews on the subject \cite{Meissner:88,Birse:96}. 
 
We believe that the discrepancies between the predictions of VMD model and experiment should rather be attributed to its approximate nature than to the idea of universality that lies behind it. In particular, we argue that the above problem can be solved by going beyond the tree-level approximation, namely, by considering the one-loop meson diagrams accounting for $1/N_c$ corrections to a tree-level result. Thus, we assume (and it will be argued in the following) that the $1/N_c$ expansion \cite{Hooft:74a,Hooft:74b,Witten:79}, where $N_c$ is a number of colors, is a relevant approximation to the issue. 

Indeed, it is known that a tree-level $V\to\gamma\to\l^+l^-$ decay amplitude, hadronic part of which is described by the matrix element $\langle 0|J|V\rangle$ of the vector quark current $J_\mu=\bar q\gamma_\mu q$, is of order $\sqrt N_c$ \cite{Witten:79}. The one-loop meson corrections are suppressed by a factor $1/N_c$. So, the $V\to \pi^+\pi^- \to \gamma\to l^+l^-$ amplitude is of order $1/\sqrt N_c$. This is precisely the order of the tree-level diagram describing the $\rho\to\pi\pi$ decay. Thus, the tree-level $\rho\to\pi\pi$ amplitude has the same weight as the $\rho\to l^+l^-$ amplitude obtained with allowance for meson loops. From that we conclude that the coupling constant $f_\rho$ extracted from formula (\ref{VLL}) should not be directly compared with $g_{\rho\pi\pi}$. Before such comparison one should take into account that $f_\rho$ contains both the tree-level and one-meson-loop contributions. Only after separating them one can judge if the hypothesis of universality is fulfilled.    

Further, from the set of all $1/\sqrt N_c$ order one-loop amplitudes contributing to $V\to l^+l^-$ decay one may exclude diagrams of the vector meson self-energy type. These diagrams were considered in \cite{Weise:96} with the conclusion that the coupling constant $f_V$ remains unaltered. The argument is as follows. Expanding the self-energy contribution $\Pi_{\mu\nu}(p^2)=g_{\mu\nu} \Pi (p^2)+\mbox{longitudinal part}$ around the physical mass of vector state $m_V$ 
\begin{equation}
\label{se}
\Pi (p^2)= \Pi (m_V^2) + (p^2-m_V^2) \frac{\partial \Pi }{\partial p^2}\vert_{p^2=m_V^2} 
+\ldots\,, 
\end{equation}
where $p$ is the four-momentum carried by the vector meson, one can require that  $\frac{\partial}{\partial p^2}\mbox{Re} \Pi\vert_{p^2=m_V^2}=0$. Then, the first term $\Pi (m_V^2)$ will determine the physical mass of the vector state through the equation $m_V^2=\mbox{\r m}_V^2+\mbox{Re} \Pi (m_V^2)$, where $\mbox{\r m}_V$ is a "bare" mass of the vector meson, which it would have if one-meson-loop contributions to its self-energy were turned off. Thus, assuming that the meson masses correspond to their empirical values, we implicitly take into account all self-energy contributions. In the following we assume this. Since we deal with the on-mass-shell vector mesons, terms of higher degrees in $(p^2-m_V^2)$ in (\ref{se}) vanish, as does the entire longitudinal part of the self-energy diagram. As a consequence, only the one-loop diagrams of two types remain, namely $V\to\pi^+\pi^-\to\gamma$ and $V\to K^+K^-\to\gamma$. 

Neglecting the imaginary part of the self-energy vector meson diagrams we neglect the relatively large width of the $\rho$ meson. The finite width effects are relevant for the electromagnetic form factors of mesons, and are not important for the rho-meson decays. This point has been addressed in \cite{Sakurai:68}, where the authors, assuming the universality condition $g_\rho =g_{\rho\pi\pi}$, derived the electromagnetic form factor of the pion with the finite width effect of the $\rho$-meson. Its application to an optimal description of the unified BABAR-BESIII data \cite{Lees:12,Ablikin:16} at the elastic region leads to the value of $\Gamma_\rho =(126.51 \pm 0.13)\,\mbox{MeV}$ \cite{Bartos:17} that differs by as much as 15\% from experimental data on the $\rho$-width. This indicates that the finite width is not only effect which should be taken into account. Indeed, as it has been recently reported \cite{Achasov:11}, the $\pi^+\pi^-$ and $K\bar K$ loops yield a consistent ground to a description of the latest experimental data on the production of the $\pi^+\pi^-$  pair in $e^+e^-$ annihilation at $\sqrt{s} < 1\, \mbox{GeV}$. Notice that when considering the pion form factor, meson loops correct the extracted value of the $\rho$ width to its phenomenological value. As we will show below, in the case of $V\to l^+l^-$ decays, meson loops $V\to \Phi\Phi\to \gamma$ lead to the correct description of two-lepton decay modes with the original value of the coupling constant $g_\rho =6.0$. 

The importance of taking one-loop meson diagrams into account when describing the electromagnetic form factor of the pion was also noted in \cite{Weise:96}. Importantly, the dominant role of the kaon loop in the electromagnetic decays of scalar mesons $f_0(980)\to \omega (\rho)\gamma$ has been proved by calculations made in \cite{Kuraev:09}. It should be stressed that in spite of the great work done on the study of the pion form factor, the role of meson loops in the $V\to l^+l^-$ decays has not yet been addressed. 

Our consideration is based on the effective Lagrangian of the extended $SU(3)\times SU(3)$ chiral symmetric Nambu-Jona-Lasinio (NJL) model \cite{Nambu:61a,Nambu:61b,Volkov:86,Ebert:82,Ebert:83,Volkov:84,Volkov:86,Ebert:86,Osipov:92,Volkov:06}, where the $\rho\to\pi\pi$ decay coupling constant $g_{\rho\pi\pi}=g_\rho=6.0$ is one of the main input parameters of the theory. This symmetry is spontaneously and explicitly broken to its isospin $SU(2)$ subgroup. The electromagnetic interactions are chosen to have a gauge invariant form  
\begin{equation}  
\label{Vg}
{\cal L}_{\gamma V} =\frac{e}{2} F^{\mu\nu}\sum_V \frac{V_{\mu\nu}}{f_V} +e \mathcal A^\mu J^{(mes)}_\mu,
\end{equation}
where $V_{\mu\nu}=\partial_\mu V_\nu -\partial_\nu V_\mu$, $F_{\mu\nu}=\partial_\mu \mathcal A_\nu -\partial_\nu \mathcal A_\mu$, $\mathcal A_\mu$ is an electromagnetic field, and $J^{(mes)}_\mu$ is electromagnetic current of mesons (see eq.(\ref{emcur}) for details). Otherwise, the theory will not have the direct $\pi\pi\gamma$ or $KK\gamma$ couplings. This form proves to be significantly better than the standard VMD form without derivatives in the description of the electromagnetic form factor of the pion \cite{Weise:96}. 

In the case of an exact $SU(3)$ flavor symmetry, $m_\rho=m_\omega =m_\phi$, there is only one independent coupling constant $f_V$, e.g., $g_\rho =6.0$. Two other constants can be expressed through $g_\rho$ as it follows $f_\omega =3g_\rho =18.0,  f_\phi =-\frac{3}{\sqrt 2} g_\rho =-12.7 $. Using a phenomenological value of $m_\rho =775.26 \pm 0.25\, \mbox{MeV}$, one finds from (\ref{VLL})      
\begin{eqnarray}
\label{dw1}
\Gamma_{\rho\to e^+e^-}&=&  4.8\, \mbox{keV}, \quad (7.04 \pm 0.06 \, \mbox{keV}), \\ 
\label{dw2}
\Gamma_{\omega\to e^+e^-}&=&  0.53\, \mbox{keV}, \quad (0.60 \pm 0.02 \, \mbox{keV}),\\
\label{dw3}
\Gamma_{\phi\to e^+e^-}&=& 1.07\, \mbox{keV}, \quad (1.26 \pm 0.01\, \mbox{keV}),    
\end{eqnarray}
where in the parentheses the corresponding experimental data are shown \cite{PDG:20}. These simple estimates show qualitatively how the $SU(3)$ symmetry, VMD and universality work in the absence of the one-loop contributions. 

The material of the paper is distributed as follows. In Section II, we discuss the relationship between dynamic symmetries and universality. In Section III, we consider how universality arises in chiral theories. The important role of the $1/N_c$ expansion for realistic theories is emphasized. These two sections are introductory and help to understand the essence of further calculations. In Section IV, the contribution of one-loop quark diagrams is calculated, and, in Section V, calculations of one-loop meson diagrams are presented. Here the main results of our calculations are contained. The Section VI summarizes the results.

\section{Dynamic symmetries and universality}
The idea of universality is associated with dynamic symmetries. These are specified either by local gauge groups or by nonlinear and inhomogeneous realizations of algebraic symmetries. As is known, dynamic symmetries not only determine the form of the Lagrangian, but also yield the low-energy theorems for massless (or very light) bosons. These theorems impose restrictions on the response of physical systems to slowly varying gauge fields and, as a result, determine the universality of the interaction in the sense that the first orders of the expansion in the coupling constant begin to coincide with the first orders of the expansion in powers of energy. This makes it possible, regardless of the value of the coupling constant, to use the corresponding effective field theory to obtain reasonable results in the low-energy region. It is important that any theory having the same low-energy spectrum of particles will have the same answer for the leading corrections independent of what the high-energy completion might turn out to be. 

To illustrate above statements let us give the classic example from the quantum electrodynamics (QED) \cite{Euler:36}. In QED, at very low energies of photons $\omega\ll m_e$, where $m_e$ is the electron mass, one can integrate out the non essential fermion degrees of freedom obtaining the effective Lagrangian for constant electromagnetic field   
\begin{equation}
\label{EH}
\mathcal L_{\gamma}=-\frac{1}{4}F^2+\frac{\alpha^2}{360m_e^2} 
\left[4F^4+ 7\left(F\tilde F \right)^2\right] + \ldots\, ,
\end{equation}
where the following notations are used  $F^2=F^{\mu\nu}F_{\mu\nu}$, $F\tilde F= F^{\mu\nu}\tilde F_{\mu\nu}$, $\tilde F^{\mu\nu}=1/2\, e^{\mu\nu\rho\sigma} F_{\rho\sigma}$. Since the photon energy is small, the tensor of electromagnetic field is slowly varying. From (\ref{EH}), one reads off that the effective Lagrangian is an  expansion in powers of derivatives and coupling constant $\alpha$. The quantum correction to the leading term is suppressed both by powers of $(\omega /m_e)^4$ and $\alpha^2\sim e^4$, i.e., the expansions in energy and in coupling constant coincide. It is easy to understand that the structure of the expansion is a consequence of the covariance of the theory and its gauge symmetry. Due to covariance, the result depends on the tensor of the electromagnetic field, and the gauge nature of the photon-electron interaction makes the degree of charge to be always equal to the degree of the 4-potential $\mathcal A_\mu$. This result is universal: if the problem was considered in which the electromagnetic field would interact with vacuum fluctuations of proton-antiproton pairs, then one would get the same formula, but with a replacement $m_e\to m_p$. Moreover, it is not at all necessary that the charged pairs be fermions. The result will remain the same for bosons. This becomes especially clear when using the proper-time method of Fock and Schwinger \cite{Schwinger:51}, where, even in the fermionic case, a transition is made from the Dirac operator to its quadratic form - the second-order Klein-Gordon differential operator to obtain (\ref{EH}). 

The universality of electromagnetic interactions can also be formulated as a statement about the non-renormalizability of the electric charge by the effects of strong interactions. It is a simple consequence of conservation of the currents and some commutation properties of charges, defined by these conserved currents \cite{DeAlfaro:73}.

\section{Large $N_c$ limit and universality}
In thinking about the low-energy regime of QCD, at the first stage one can neglect the small bare masses of the up, down and strange quarks. In this case, the left- and right-handed quarks do not interact with each other and the whole theory admits an $U(3)_L\times U(3)_R$ chiral symmetry. The $U(1)_A$ axial anomaly reduces this to the $SU(3)_L\times SU(3)_R\times U(1)_{L+R}$ group. However, the chiral symmetry is not realized in the Wigner-Weyl mode, the ground state is asymmetric under $SU(3)_L\times SU(3)_R$ \cite{Nambu:61a,Nambu:61b}. The result is that chiral symmetry is spontaneously broken down to the vectorial subgroup of flavor and hypercharge, generated by the vector currents of $SU(3)_{L+R} \times U(1)_{L+R}$ group. In accord with Goldstone theorem \cite{Goldstone:61} the spectrum of massless QCD must therefore contain $N_f^2-1 = 8$ massless pseudoscalar bosons. 

Goldstone particles arise in systems with a degenerate vacuum, the symmetry of which is lower than the symmetry of the original Lagrangian. Qualitatively, this can be interpreted as a reaction of the system aimed at restoring the lost algebraic symmetry. In this case, the transformation properties of Goldstone particles are determined by nonlinear transformations, which, together with the requirement of invariance of the effective Lagrangian with respect to these chiral transformations, correspond to a symmetry of dynamic type \cite{Weinberg:68}.  

Of course, chiral symmetry is approximate, because in the real world the quark masses are not exactly zero. This gives rise to an explicit chiral symmetry breaking effects. As a result, the low-energy theorems following from the chiral dynamics for Goldstone particles are not exactly correspond to the nature: a careful study of the explicit symmetry breaking effects is required. This is a subject of the chiral perturbation theory \cite{Gasser:84,Gasser:85a,Gasser:85b}.  

Another distinction of chiral dynamics is the non-fundamental character of pseudo-Goldstone states. Their quark - antiquark structure significantly distinguishes these states from fundamental gravitational or electromagnetic fields. 

That mesons and nucleons are not fundamental states was particularly emphasized by Sakurai \cite{Sakurai:69,Sakurai:60} when formulating his idea of the universality of vector mesons. How does this circumstance affect the predictions based on the idea of universality of strong interactions? We suppose that this question can be answered within the framework of $1/N_c$ expansion. In various ways this expansion is reminiscent of known phenomenology of hadron physics, indicating that an expansion in powers of $1/ N_c$ may be a good approximation even at $1/ N_c=1/3$. This conclusion can also be reached on the basis of the QCD sum rules, provided the channels considered correspond to modest values of the critical mass $M_{crit}^2$ \cite{Shifman:81}. For light mesons and quarks $M_{crit}^2\sim 2\,\mbox{GeV}^2$. It will be argued here that it is from the point of view of the $1/N_c$ expansion the application of universality to the chiral dynamics makes the most sense. 

To spice up our discussion, it is useful to turn to the theory with local four-quark interactions of the NJL type, which provides a general framework for constructing both linear and nonlinear effective meson Lagrangians describing the dynamics of composite quark-antiquark states. Let us consider the model with the Lagrange density \cite{Osipov:00}
\begin{eqnarray}
\label{lag}
{\cal L}&=&\bar q(i\gamma^\mu\partial_\mu -\mathcal M)q + \frac{G_S}{2}\left[(\bar qq)^2+(\bar qi\gamma_5\vec\tau q)^2 \right] \nonumber \\
&-& \frac{G_V}{2}\left[(\bar q\gamma^\mu\vec\tau q)^2+(\bar q\gamma^\mu\gamma_5\vec\tau q)^2 \right].
\end{eqnarray}
Here $q$ is an isodoublet spinor of light quarks $\bar q=(\bar u, \bar d)$; summation over color indices is implicit, $\vec \tau$ are the isospin Pauli matrices; coupling constants $G_S$ and $G_V$ have the same dimension $[G_S]=[G_V]=M^{-2}$; a current quark mass matrix $\mathcal{M}= \mbox{diag}(\hat m_u, \hat m_d)$ is chosen in a form ($\hat m_u=\hat m_d\equiv \hat m$) preserving the isospin symmetry. 

Without a quark mass term ${\cal L}$ would possess a continuous $SU(2)_V \times SU(2)_A$ symmetry. The global transformations of this group can be parameterized by six real parameters: $\alpha_i$ and $\beta_i$, where $i=1,2,3$. The infinitesimal transformation of quark fields $\delta q=q'-q$ is 
\begin{eqnarray}
\label{qt}
&&\delta q=i(\alpha +\gamma_5\beta )q, \quad \delta \bar q=i\bar q(-\alpha+\gamma_5\beta ), \nonumber\\ &&\alpha=\alpha_i \frac{\tau_i }{2} , \quad \beta=\beta_i  \frac{\tau_i}{2} .
\end{eqnarray}
Notice that $SU(2)_V \times SU(2)_A$ symmetry is realized here as a homogeneous linear transformation that includes the standard isospin transformation $\alpha$ and the chiral transformation $\gamma_5\beta$. 

The description of collective modes can be facilitated if one introduces the new Lagrangian with the same dynamical content 
\begin{eqnarray}
\label{smat1}
&&Z=\!\!\int\!\!{\cal D}q{\cal D}\bar q{\cal D}s{\cal D}\vec p{\cal D}\vec v_\mu {\cal D}\vec a_\mu  \nonumber \\
&&\exp i\!\!\int\!\! d^4x\left[\bar qDq -\frac{(s+\hat m)^2+\vec p^{\,2}}{2G_S} +\frac{\vec v_\mu^{\,2}+\vec a_\mu^{\,2}}{2G_V} \right].
\end{eqnarray}
Here $D$ is the Dirac operator in the external boson fields 
\begin{equation}
\label{Dirac}
D = i\gamma^\mu\partial_\mu +s+i\gamma_5 p +\gamma^\mu v_\mu + \gamma^\mu\gamma_5 a_\mu,
\end{equation}
where $s=s \cdot 1,\, p=\vec p\,\vec\tau,\, v_\mu =\vec v_\mu\vec\tau,\, a_\mu =\vec a_\mu\vec\tau$ are scalar, pseudoscalar, vector and axial-vector auxiliary fields. 

The infinitesimal action of the chiral group on quarks (\ref{qt}) helps to obtain a transformation law for quark-antiquark auxiliary fields
\begin{eqnarray}
\label{sptlaw1}
\delta s&=&\{\beta, p\}, \nonumber\\
\delta p&=&i[\alpha, p]-\{\beta, s\}, \nonumber \\
\delta v_\mu &=&i[\alpha, v_\mu ]+i[\beta, a_\mu], \nonumber \\ 
\delta a_\mu &=& i[\alpha, a_\mu ]+i[\beta, v_\mu].
\end{eqnarray}

Spontaneous symmetry breaking ($s=\sigma -m, \langle\sigma\rangle =0$) leads to the mixing between pseudoscalar and axial-vector fields already in the one-quark-loop approximation, i.e. in the same order at which the effective potential develops the non-symmetric ground state. To avoid the mixing term one should define a new axial-vector field $ A_\mu$
\begin{equation}    
\label{mix}
a_\mu=A_\mu +\kappa m\partial_\mu p. 
\end{equation}
Constant $\kappa$ must be fixed to avoid $p A_\mu$ mixing. The replacement (\ref{mix}) adds to the Lagrange density (\ref{smat1}) a Yukawa-type vertex $\kappa m\,\bar q\gamma^\mu \gamma_5 \vec\tau q\,\partial_\mu\vec p$ known from the theory of pion-nucleon forces.  

Integrating over the quark fields in (\ref{smat1}), and keeping only the leading terms in the inverse constituent quark mass expansion of the corresponding heat kernel (it is assumed that the Fock-Schwinger proper time technique \cite{Schwinger:51} is applied to generate $1/m^2$ expansion), one arrives to the effective Lagrange density describing the strong dynamics of collective boson fields. 
\begin{eqnarray}
\label{linv}
&&\mathcal L'=-\frac{\hat m\,\mbox{tr}\, (\sigma^2+p^{2})}{4mG_S} +\frac{\mbox{tr}\,\left[v_\mu^{\,2}+(A_\mu+\kappa m\partial_\mu p\,)^2\right] }{4G_V}\nonumber\\
&&+I_2(m^2)\, \mbox{tr}\left\{(\bigtriangledown_\mu\sigma )^2 + (\bigtriangledown_\mu p)^2 
-(\sigma^2-2m\sigma +p^2)^2 \right.\nonumber\\
&&\left.\ \ \ \ \ \ \ \ \ \ \ \ \ \ \ -\frac{1}{3}(v_{\mu\nu}^2+A_{\mu\nu}^2) \right\}.  
\end{eqnarray}
Here the trace is taken over isospin matrices. The main elements of (\ref{linv}) are $SU(2)\times SU(2)$ covariant objects
\begin{eqnarray}
\label{covder}
\bigtriangledown_\mu\sigma &=&\partial_\mu\sigma -\{a_\mu, p \}, \nonumber\\
\bigtriangledown_\mu p &=&\partial_\mu p-i[v_\mu, p]+\{a_\mu, \sigma -m\}, \nonumber \\
v_{\mu\nu}&=&\partial_\mu v_\nu - \partial_\nu v_\mu -i[v_\mu , v_\nu ] -i[a_\mu , a_\nu ], \nonumber \\
A_{\mu\nu}&=&\partial_\mu A_\nu - \partial_\nu A_\mu -i[v_\mu, a_\nu  ] -i[a_\mu , v_\nu].
\end{eqnarray}
The regularized mass-dependent factor $I_2(m^2)$ is
\begin{equation}
\label{j01}
I_2(m^2)=\frac{N_c}{16\pi^2}\left[\ln\left(1+\frac{\Lambda^2}{m^2}\right)-\frac{\Lambda^2}{\Lambda^2+m^2}\right], 
\end{equation}
where $\Lambda$ is a covariant cutoff of the corresponding one-quark-loop integral.

The gap equation arises in the form of the condition which cancels a $\sigma$-tadpole contribution in  (\ref{linv}). This equation accounts for contributions of quarks of different flavours $N_f=2$ and colors $N_c=3$, as well as the current quark mass $\hat m$
\begin{equation}
\label{gap}
\frac{4\pi^2}{N_fN_c\Lambda^2 G_S}\left(1-\frac{\hat m}{m}\right)= 1-\frac{m^2}{\Lambda^2}\ln\left(1+\frac{\Lambda^2}{m^2}\right). 
\end{equation}
The Lagrangian density $\mathcal L'$ does not contain $p A_\mu$-mixing. This is because of the cancellation which occurs between three different contributions to the nondiagonal $pA_\mu$-mixing term in $\mathcal L'$. It gives the numerical value of $\kappa$ 
\begin{equation}
\label{kappa}
\frac{1}{2\kappa}=m^2+ \frac{1}{16G_V I_2(m^2)}.
\end{equation}

The free part of the Lagrangian $\mathcal L'$ will have a canonical form after the rescaling of meson fields
\begin{equation}
\label{coupling}
\sigma =g_\sigma \bar\sigma, \quad \vec p=g_\pi\vec\pi, \quad \vec v_\mu=\frac{g_\rho}{2}\vec\rho_\mu, \quad \vec A_\mu =\frac{g_\rho}{2}\vec a_{1\mu}. 
\end{equation}
As a result, we have 
\begin{eqnarray}
\label{spectr}
&&g_\sigma^2=\frac{1}{4I_2 (m^2)},\quad  g_\pi^2=Zg_\sigma^2,\quad  g_\rho^2=6g_\sigma^2, \nonumber \\ 
&& m_\pi^2=\frac{\hat mg_\pi^2}{mG_S}, \quad m_{\bar \sigma}^2=4m^2+Z^{-1}m_\pi^2, \nonumber \\
&& m_\rho^2=\frac{3}{8G_V I_2(m^2)},\quad m_{a_1}^2=m_\rho^2+6m^2,   
\end{eqnarray}
where $Z^{-1}=1-2\kappa m^2=g_A$.

What one can learn from this simple example? Let us first indicate that as it follows from formulas (\ref{coupling}) and (\ref{spectr}) there is a precise relation between coupling constants $g_\sigma, g_\pi, g_\rho$.  

Next, let us consider the vector mesons. As can be verified from (\ref{linv}), the following equalities hold
\begin{equation}
\label{univS}
g_\rho = g_{\rho\pi\pi}  = g_{\rho\rho\rho} = g_{\rho qq}.
\end{equation}
These are the well-known universality relations of Sakurai \cite{Sakurai:69} (up to the last term, where the rho-quark-quark coupling replaces the rho-nucleon-nucleon coupling). 

As one of the arguments in favor of universality relations (\ref{univS}), a hypothesis was put forward about the invariance of the interaction Lagrangian under the local action of the group of the isospin symmetry. The role of gauge fields was assigned to an isotriplet of vector rho-mesons \cite{Sakurai:69}. Indeed, four-quark interactions in (\ref{lag}) possess a higher symmetry. They are invariant under local $SU(2)_V \times SU(2)_A$ transformations. It is easy to see that the term $\bar qDq$ in (\ref{smat1}) is also symmetric under the action of this group. Thus, dynamic symmetry (if one neglects the $\rho$-meson mass term) is hidden behind the relations (\ref{univS}), which makes them more significant.

Despite the fact that mesons in the NJL model are composite quark-antiquark states, in the Lagrangian (\ref{linv}) them correspond the local fields. Of course, if one would take into account the higher order terms (in derivative expansion of the effective action) neglected in (\ref{linv}), one would reveal the composite structure of meson states. In momentum space, this would lead to the momentum dependence of the coupling constants in (\ref{univS}), i.e., $g_\rho\to g_\rho (q^2)$. The corresponding technique has been developed in \cite{Osipov:94,Osipov:96}. For the slowly changing form factor, $g_\rho (q^2)$, the principle of universality may still have sense even for the on-shell coupling constants. On the other hand, the large $N_c$ limit teaches us that meson physics at $N_c=\infty$ is described by the tree diagrams of an effective local Lagrangian with the local meson fields. If we adhere to this point of view (supposing that $N_c=3$ is a good approximation to the $N_c=\infty$ case), then the calculation of the momentum dependence of the coupling constants based on one-quark-loop diagrams does not make sense. The point is that the higher order terms of the expansion of the effective action contribute at the same order in $1/N_c$, and therefore the model comes into the conflict with QCD at large $N_c$. In this case, one should account for meson loops instead, which are suppressed as $1/N_c$. Thus, limiting ourselves to the Lagrangian (\ref{linv}), we assume that $g_\rho$ is a constant on the interval $0<q^2<\Lambda^2$. Actually, this assumption is not new. What is new is the observation that this hypothesis is true at large $N_c$. 

The effective Lagrangian of pseudo-Goldstone particles, which we arrive at by excluding from (\ref{linv}) the scalar field $\sigma$ (see, for instance, \cite{Osipov:00}), collects all soft-pion low-energy theorems (at tree-level order). It gives us the well-known result of the nonlinear sigma model \cite{Donoghue:92}, which, due to its universality, is used as a leading approximation in the chiral perturbation theory.

\section{$SU(3)$ violation: quark-loop level}
Let us now turn to our main task. For this we will use a more advanced version of the NJL model, which contains, in addition to the up and down quarks, a heavy strange quark \cite{Volkov:93}. In the NJL model, the constituent masses of the up, down and strange quarks $m_u, m_d, m_s$ are the result of the spontaneous and explicit chiral symmetry breaking. A gap equation (\ref{gap}) illustrates how this happens. Here we consider the model with $SU(2)$ isospin symmetry $m_u=m_d\neq m_s$. The origin of the $SU(3)$ flavor symmetry violation is the nonzero values of the bare current quark masses $\hat m_u=\hat m_d \neq \hat m_s$. The model estimation of these values are $\hat m_u = \hat m_d = 3\,\mbox{MeV}$ and $\hat m_s = 90\,\mbox{MeV}$ \cite{Volkov:93}. The current quarks get heavier in the phase with broken chiral symmetry, where they are constituent quarks (with the following masses $m_u=m_d=270 \, \mbox{MeV}$, and $m_s=420 \, \mbox{MeV}$) inside the colorless mesons. 

The effective local vertices of the induced meson Lagrangian are obtained from the corresponding quark one-loop diagrams with external meson fields by the series expansion in the inverse powers of heavy quark masses. The details of the model can be found in \cite{Volkov:93}. Here we only remind the main steps which are essential for fixing parameters of the model.  

The decay process of a neutral vector meson, for instance the $\rho^0$-meson, is described by the diagram shown in Fig.\,1. 
\begin{figure}[h]
\label{fig1}
\center{\includegraphics[scale = 0.4]{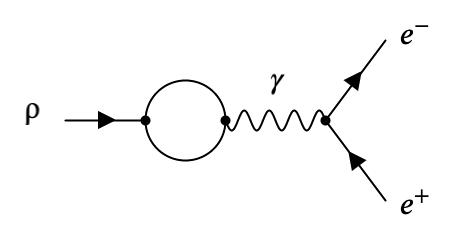}}
\caption{The tree-level diagram of $\rho \to e^+e^-$ decay. The $\rho$-meson emits the photon through the quark loop.}
\end{figure}
In the NJL model, the calculation of the $V\to\gamma$ transition through the quark loop, besides the standard photon-quark-antiquark vertex, requires the Lagrangian density describing the strong interactions of vector mesons 
\begin{eqnarray}
\mathcal L_{\bar qqV}&\Rightarrow& \frac{g_\rho}{2}\rho^0_\mu (\bar u \gamma^\mu u -\bar d \gamma^\mu d) \nonumber\\
&+& \frac{g_\omega}{2}\omega_\mu (\bar u \gamma^\mu u +\bar d \gamma^\mu d) +\frac{g_\phi}{\sqrt 2} \phi_\mu \bar s\gamma^\mu s,
\end{eqnarray}
where couplings $g_\rho, g_\omega, g_\phi$ are chosen to redefine the kinetic part of vector meson free Lagrangian to its standard form. If isospin symmetry is fulfilled, $g_\rho =g_\omega$ the calculation of $V\to\gamma$ transition of Fig.\,1 leads to the Lagrangian density (\ref{Vg}), where $f_\rho =g_\rho$, $f_\omega =3g_\rho$, and $f_\phi =-\frac{3}{\sqrt 2}g_\phi$. The coupling constants are 
\begin{equation}
\label{grho}
g_\rho =\sqrt{\frac{3}{2I_2(m_u^2)}}, \quad g_\phi =\sqrt{\frac{3}{2I_2(m_s^2)}}.
\end{equation}
The logarithmically divergent integrals $I_2(m^2_1,m^2_2)$ are regularized through the covariant cutoff $\Lambda$   
\begin{equation}
I_2(m_1^2,m^2_2) = -i\frac{N_{c}}{(2\pi)^{4}}\!\int \! \mathrm{d}^{4}k\, 
\frac{\theta (\Lambda^2+k^2)}{(m_1^{2} - k^2) (m_2^{2} - k^2)}. 
\end{equation}
The shorthand $I_2(m^2_u)$ in (\ref{grho}) is introduced for the case of equal masses $I_2(m^2_u)\equiv I_2(m^2_u,m^2_u)$. In this specific case, the expression coincides with (\ref{j01}).   

In the case of charged pions and kaons, the quark-meson vertices are described by the Lagrangian density 
 \begin{equation}
 \mathcal L_{\bar qq\Phi} \Rightarrow  \sqrt 2 \bar{u} i\gamma_5 \left( g_\pi \pi^+ d+g_K K^+ s\right) + h.c.   
 \end{equation}
The coupling constants $g_\pi$ and $g_K$ are given by 
\begin{equation}
\label{gK}
    	g_{\pi} = \sqrt{\frac{Z_{\pi}}{4 I_{2} (m_u^2)}}, \quad g_{K} = \sqrt{\frac{Z_{K}}{4 I_{2}(m^2_u, m^2_s)}}.
\end{equation}
There are two types of contributions here. Firstly, the integrals $I_2$ determine the redefinition of pseudoscalar fields to make that kinetic part of their free Lagrangian to have standard form. Secondly, due to spontaneous symmetry breaking, the partial Higgs mechanism takes place. The axial-vector -- pseudoscalar mixing term, $A_\mu \partial^\mu \Phi$, appears in the free meson Lagrangian. This non-diagonal term must be canceled by an appropriate redefinition of the longitudinal component of the massive axial-vector field $A_\mu = A'_\mu + \kappa m \partial_\mu\Phi$. The result is that the kinetic term of a free pseudoscalar field get an additional factor, which must also be compensated by the specific choice of constants $Z_\pi$ and $Z_K$. They are     
\begin{equation}
\label{Z}
Z_{\pi}^{-1} = 1 - 6\frac{m^2_u}{m^2_{a_{1}}}, \quad
Z_{K}^{-1}  = 1 - \frac{3}{2}\frac{(m_{u} + m_{s})^{2}}{m^{2}_{K_{1A}}}.
\end{equation}
The coupling $Z_\pi$ originated by the mixing between the pion and the $a_1(1260)$ mesons ($m_{a_{1}} = 1230\pm 40 $\,MeV). The factor $Z_K$ is due to the mixing between the kaon and the $K_{1}$ fields. There are two candidates, known as $K_1(1270)$ and $K_{1}(1400)$. These physical states are the mixture of $K_{1A}$ and $K_{1B}$ fields. The latters have different types of couplings with quark-antiquark pair, namely $K_{1A}\propto \gamma_5\gamma_\mu$, $K_{1B}\propto \gamma_5\partial_\mu$. The constant $m^{2}_{K_{1A}}$ is determined through the mixing angle $\alpha =57^\circ$ of these states \cite{Osipov:85,Suzuki:93}     
\begin{equation}    	
m^{2}_{K_{1A}} = \left(\frac{\sin^{2}{\alpha}}{m^{2}_{K_{1}(1270)}} - \frac{\cos^{2}{\alpha}}{m^{2}_{K_{1}(1400)}}\right)^{-1}.
\end{equation}
A general mathematical framework to deal with axial-vector--pseudoscalar mixing in the effective Lagrangians has been developed in \cite{Osipov:17}.    

Finally, let us fix the parameters of the model. The value of $g_\rho$ is determined from the $\rho\to\pi\pi$ decay, that gives $g_\rho =6.0$. Then, using eq. (\ref{grho}), one finds $I_2(m_u^2)=3/(2g_\rho^2)=0.042$. The charged pion weak decay constant $\pi^+\to\mu^+\nu_\mu$, $f_\pi =92.2\,\mbox{MeV}$, and the quark-level Goldberger-Treiman relation $g_\pi=m_u/f_\pi$ allow us to establish an equation for the non-strange quark mass
\begin{equation}  
m_u= \frac{1}{2} f_\pi \sqrt{\frac{Z_\pi}{I_2(m_u^2)}}=f_\pi g_\rho \sqrt{\frac{Z_\pi}{6}}.  
\end{equation}
This quadratic equation can be easily solved in terms of known phenomenological parameters. As a result we have
\begin{equation}
\label{mu}
m_u^2=\frac{m^2_{a_1}}{12} \left[1\pm \sqrt{1-\left(\frac{2g_\rho f_\pi}{m_{a_1}}\right)^2}\right].
\end{equation}
It allows us to obtain the value of constituent quark masses $m_u = m_d = 270\,$MeV (for the negative sign in the solution (\ref{mu})). Now from $I_2(m_u^2)$ we find the quark loop cutoff parameter $\Lambda =1265\, $MeV.  

To establish the value of strange quark mass we use the Goldberger-Treiman relation for the kaon $g_K$
\begin{equation}
g_K=\frac{m_u+m_s}{2f_K},
\end{equation} 
where $f_K\simeq 1.2 f_\pi$ \cite{Beringer:12}. Using eq. (\ref{gK}), one comes to the formula, where $m_s$ is the only unknown parameter
\begin{equation}  
f_K=(m_u+m_s)\sqrt{\frac{I_2(m_u^2, m_s^2)}{Z_K}}.
\end{equation}
Solving numerically this equation we obtain the value of the constituent strange quark mass $m_s=420$\,MeV. 
For above values of the quark masses and the cutoff $\Lambda = 1265\,\mbox{MeV}$, the condensates are $ \langle\bar u u\rangle^{1/3} =\langle\bar dd\rangle^{1/3}= - 304\,\mbox{MeV}$, $\langle\bar ss\rangle^{1/3} = - 335\,\mbox{MeV}$. These values can be compared with the lattice QCD results  $\langle \bar ll \rangle^{\overline{MS}}(2\mbox{GeV}) = (- 283(2)\,\mbox{MeV})^3$ and $\langle \bar s s\rangle^{\overline{MS}}(2\mbox{GeV}) = (- 290(15)\,\mbox{MeV})^3$, where $l$ is a light quark with mass equal to the average of the $u$ and $d$ quarks \cite{Bazavov:13}. 

It is known that the gap-quation of the NJL model leads to underestimated values of the current quark masses, which is reflected in the magnitude of the quark condensates. The latter is due to the fact that the value of the chiral condensate is related with masses of pseudoscalars through the well-known Gell-Mann, Oakes, Renner relation
\begin{equation}
m_\pi^2 f_\pi^2 = - \frac{m^0_u+m_d^0}{2} \left(\langle \bar uu\rangle + \langle \bar dd\rangle\right).
\end{equation}
So, all is arranged in such a way that model describes correctly the masses of pions and kaons (it is interesting to note that this parametrization of the model leads to a perfect agreement with the experimental value for the kaon mass, giving $m_K=494$\,MeV). 

Now, we can estimate $g_\phi =7.5$, and, as a consequence, this predicts $\Gamma_{\phi\to e^+e^-}=0.9\, \mbox{keV}$. One can see that $SU(3)$ symmetry breaking at the level of quark loops leads to an even greater disagreement with the experimental data on the $\phi\to e^+e^-$ decay mode. 

\section{$SU(3)$ violation: meson-loop level}
\begin{figure}[h]
\center{\includegraphics[scale = 0.45]{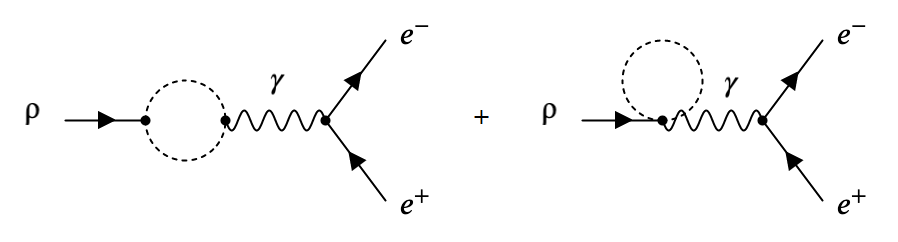}}
\caption{Diagrams describing contributions from meson loops to $\rho\to e^+ e^-$ decay. The dotted line indicates pions or kaons.}
\label{Intermediate}
\end{figure}

Let us consider now the contributions to the amplitudes due to the meson loops. The corresponding diagrams for the $\rho\to e^{+} e^{-}$ decay are shown in Fig.\,\ref {Intermediate}. 

The vertices, describing $V\to\Phi\Phi$ decays in the case of exact $SU(3)$ symmetry are given by the Lagrangian density
\begin{eqnarray}
\label{rpp}
&&{\cal L}_{V\Phi\Phi}=-i\frac{g_\rho}{4}\,\mbox{tr}\left(V_\mu [\Phi , \partial^\mu\Phi ] \right)\nonumber \\
&&=-i\frac{g_\rho}{2}\left[ \rho^0_\mu \left( 2\pi^+ \! \stackrel{\leftrightarrow}{\partial^\mu}\!  \pi^- 
      + K^+ \! \stackrel{\leftrightarrow}{\partial^\mu}\! K^- 
      + \bar K^0 \! \stackrel{\leftrightarrow}{\partial^\mu}\! K^0 \right) \right. \nonumber \\
&&+\left. (\omega_\mu -\sqrt 2 \phi_\mu ) (K^+ \! \stackrel{\leftrightarrow}{\partial^\mu}\! K^- 
      - \bar K^0 \! \stackrel{\leftrightarrow}{\partial^\mu}\! K^0) \right].   
\end{eqnarray}
The left-right derivative is $\phi_1 \! \stackrel{\leftrightarrow}{\partial^\mu}\!  \phi_2 = \phi_1\partial^\mu \phi_2 - (\partial^\mu \phi_1) \phi_2$, and nonets of the pseudoscalar $\Phi$ and vector $V$ fields are given by the following matrices 
\begin{eqnarray}
\label{Ps}
\Phi &=&\left(\begin{array}{ccc}
\pi^0  & \sqrt 2 \pi^+ & \sqrt 2 K^{+} \\
\sqrt 2 \pi^- & - \pi^0 & \sqrt 2 K^0   \\
\sqrt 2 K^-  & \sqrt 2 \bar K^0 & 0
\end{array}\right) +\sum_{a=0,8}\eta_a\lambda_a, \\
\label{Vm}
V &=&\left(\begin{array}{ccc}
\omega +\rho^0  & \sqrt 2 \rho^+ & \sqrt 2 K^{*+} \\
\sqrt 2 \rho^- & \omega - \rho^0 & \sqrt 2 K^{*0}   \\
\sqrt 2 K^{*-}  & \sqrt 2 \bar K^{*0} & \sqrt 2 \phi
\end{array}\right),
\end{eqnarray}
where the last term in (\ref{Ps}) contains neutral singlet and octet components $\eta_0$ and $\eta_8$, the superposition of which describes physical $\eta, \eta'$ states; $\lambda_0=\sqrt{2/3}\,I$, with $I$ the $3\times 3$ identity, and $\lambda_8 =1/\sqrt 3\,\mbox{diag} (1,1,-2)$. We assume in (\ref{Vm}) that vector $\rho^0, \omega, \phi$ mesons have the following quark structure: $\rho^0=(u\bar u-d\bar d)/\sqrt 2$, $\omega =(u\bar u+d\bar d)/\sqrt 2$, and $\phi =s\bar s$ consists only of strange quarks. Therefore, in the following we neglect the small $\phi -\omega$ mixing.

To take into account the effects of $SU(3)$ symmetry breaking, we, as in the previous section, should consider the quark triangle diagrams leading to the effective Lagrangian (\ref{rpp}) by taking into account the strange quark effects. The only consequence of such calculations is that one should replace the coupling $g_\rho\to g_\phi$ in the $\phi \bar KK$ vertex of (\ref{rpp}).         

The $U(1)$ gauge invariant electromagnetic interactions of the pseudoscalar $\Phi$ mesons in the NJL model are given by the effective Lagrangian density 
\begin{eqnarray}
\label{emcur}
\mathcal L_{\gamma\Phi\Phi}&=&\frac{ie}{2}\mathcal A^\mu \mbox{tr}\left(Q [\partial_\mu\Phi, \Phi ] \right) \nonumber \\
&=& -ie \mathcal A^\mu \left( \pi^+ \! \stackrel{\leftrightarrow}{\partial_\mu}\!  \pi^-  
 + K^+ \! \stackrel{\leftrightarrow}{\partial_\mu}\!  K^- \right), 
\end{eqnarray}
where $Q=\mbox{diag}(2/3, -1/3, -1/3)$ is the quark charge matrix in relative units of the proton electric charge $e>0$. This is the usual form which can be obtained independently by the standard $U(1)$ gauging of the free meson Lagrangian. 

Using these Lagrangians in the calculations of diagrams in Fig.\,\ref{Intermediate} one obtains the following amplitudes
\begin{eqnarray}
        T_{\rho \to e^+e^-}\!&=& \frac{e^2}{g_{\rho}}\!\left[1+ \frac{g^2_{\rho}}{6m^2_{\rho}}\left(2 I_{\rho\pi} + I_{\rho K}\right)\right]\! \epsilon_{\mu}(p) l^{\mu}, \\ 
        T_{\omega \to e^+e^-}\!&=& \frac{e^2}{3 g_{\rho}}\!\left(1+ \frac{g^2_{\rho}}{2m^2_{\omega}}I_{\omega K} \right)\!\epsilon_{\mu}(p) l^{\mu}, \\ 
        T_{\phi \to e^+e^-}\!&=&\frac{\sqrt{2}}{3}\frac{e^2}{g_{\phi}}\left(1+ \frac{g^2_{\phi}}{2m^2_{\phi}}I_{\phi K}\right)\!\epsilon_{\mu}(p) l^{\mu}.
\end{eqnarray}	
Here the first term in the parentheses represents the tree-level contribution of order $\sqrt N_c$, the other terms accounts for $1/\sqrt N_c$ corrections. We also assume that $m_\rho =m_\omega$ ($SU(2)$ symmetry), $\epsilon_\mu (p)$ is a polarization 4-vector of the corresponding vector meson with 4-momentum $p$, $l^\mu$ is a vector lepton current $l^\mu =\bar e\gamma^\mu e$. The integral over the pion-loop for the decay $\rho\to e^+ e^-$ after regularization (the covariant cutoff $\Lambda_\pi$ is introduced) has the form    
\begin{eqnarray}
\label{L1}
&&I_{\rho\pi}(m_\rho^2)=\frac{1}{16\pi^2}\left[2\Lambda_\pi^2+(m_\rho^2-6m_\pi^2)\ln\left(1+\frac{\Lambda_\pi^2}{m_\pi^2}\right)   \right. \nonumber \\
&+& 2(2\Lambda_\pi^2+m_\rho^2-4m_\pi^2 ) D(\Lambda^2_\pi) \arctan\frac{1}{D(\Lambda^2_\pi)}    
\nonumber \\
&+& \left. 2m_\rho^2 D(0)^3 \arctan\frac{1}{D(0)}\right]  - I_{tp}(m^2_\pi ,\Lambda^2_\pi),
\end{eqnarray}
where
\begin{equation}
D(\Lambda^2_\pi )=\sqrt{\frac{4(\Lambda^2_\pi +m^2_\pi)}{m^2_\rho}-1},
\end{equation}
and the last integral in (\ref{L1}) represents the contribution of the tadpole diagram, shown in Fig.\,\ref{Intermediate}
\begin{equation}
I_{tp}(m^2_\pi,\Lambda^2_\pi )= \frac{3}{8\pi^2} \left[ \Lambda_\pi^2 -m_\pi^2 \ln\left(1+\frac{\Lambda_\pi^2}{m_\pi^2}\right)\right].
\end{equation}
If expression (\ref{L1}) is expanded in a series in terms of the square of the mass of the $\rho$-meson, then such an expansion begins with a term that is completely canceled out by the contribution of the tadpole $I_{tp}(m^2_\pi,\Lambda^2_\pi )$. Thus, the integral $I_{\rho\pi}$ contains only a logarithmically divergent (at $\Lambda_\pi\to\infty$) part. The other consequence is that $I_{\rho\pi}(0)=0$ at fixed $\Lambda_\pi$.     

The integrals $I_{\rho K}, I_{\omega K}$, and $I_{\phi K}$ are obtained from $I_{\rho\pi}$ by replacing the masses $m_\pi \to m_K$, $m_\rho\to m_\omega, m_\phi$ and the cutoff  $\Lambda_\pi\to \Lambda_K$ correspondingly. The cutoff $\Lambda_K =750\,\mbox{MeV}$ is fixed from the decay width $\phi\to e^+e^-$. Then, the cutoff parameter $\Lambda_\pi = 850\,\mbox{MeV}$ is fixed from the experimental width of the processes $\rho\to e^+ e^-$. As a result, the process $\omega\to e^+e^-$ and three $\mu^+\mu^-$ decay modes have no arbitrary parameters, and the theoretical widths for all of them agree with the experimental data. The results are shown in the Table\,I. 

\begin{table}[t]
\label{Tabddd}
\begin{center}
\caption{Decay widths of neutral vector mesons into lepton pairs (in keV). The tree-level NJL results are given in the second column. The full results which take into account the one-meson-loop corrections are shown in the third column.}
\begin{tabular}{cccc}
\hline
\hline
\  Modes &\ \ \ \  NJL &\ \ \  NJL+mes. loops &\ \ \  Experiment \cite{PDG:20}  \\ 
\hline
\ \	$\rho \to e^+e^-$ &\ \  4.8 & 6.99 & 6.98 $\pm$ 0.12 \\
\ \	$\omega \to e^+e^-$ &\ \ 0.53 & 0.62 & 0.62 $\pm$ 0.02 \\
\ \	$\phi \to e^+e^-$ &\ \ 0.90 & 1.26& 1.26 $\pm$ 0.02 \\
\ \	$\rho \to \mu^+\mu^-$ &\ \ 4.28 & 6.22 & 6.72 $\pm$ 0.46 \\
\ \	$\omega \to \mu^+\mu^-$ &\ \ 0.48 & 0.56 & 0.63 $\pm$ 0.16 \\
\ \	$\phi \to \mu^+\mu^-$ &\ \ 0.84 & 1.18 & 1.22 $\pm$ 0.08 \\
\hline 
\hline
\end{tabular}
\end{center}
\end{table}

The first column shows the modes of two-lepton decays of neutral vector mesons, and the second contains estimates made on the bases of the tree approximation ($\sqrt N_c$-order). One can see that the tree level result underestimates the decay widths considered. It corresponds the following values of coupling constants: $f_\rho =6, f_\omega=18, f_\phi=-15.9$. Notice, that in the case of decays of $\rho$ and $\omega$ mesons, we deal with the case of unbroken $SU(3)$ symmetry, and for the $\phi$ meson our result takes into account the $SU(3)$ breaking effect arising from the heavier quark mass of the strange quark, $m_s>m_u$. The numbers presented in the third column include the one-meson-loop effects. They correct the results to their phenomenological values. The decay rates of similar decays of neutral vector mesons into a muon pair are also in satisfactory agreement with experiment.

It seems necessary to discuss in more detail the use of the cutoff $\Lambda_\Phi$ in the integrals $I_{V\Phi}$. Usually, in such cases, dispersion methods are used \cite{Weise:96, Achasov:11}. Our remark is that since the vector mesons are on the mass-shell, the two methods are equivalent. Indeed, the dispersion approach for the meson loop gives the following result for the integral with two subtractions 
\begin{equation}
I_{V\Phi} (p^2) = c_0+c_\Phi p^2 + \frac{p^4}{\pi}\!\! \int\limits_{4m_\Phi^2}^\infty \!\! ds \,\frac{\mbox{Im}I_{V\Phi} (s) }{s(s-p^2-i\epsilon )}, 
\end{equation}
where the subtraction constants $c_0$ and $c_\Phi$ need to be determined. Due to gauge symmetry $c_0$ should vanish. In our case $c_0$ is also vanish because of a tadpole contribution. It is clear now that on the mass shell $p^2=m_V^2$ the value of this integral depends on the only free parameter $c_\Phi$. Therefore, one can always establish the one-to-one correspondence between $c_\Phi$ and cutoff $\Lambda_\Phi$ by equating the integrals.    

\section{Conclusions}
We have shown that one-loop meson diagrams with the emission of the photon, i.e., the $V\to \pi^+\pi^-\to \gamma$ and $V\to K^+K^-\to \gamma$ transitions, can be an efficient mechanism for linking the hypothesis of vector meson universality with the empirical data on the electromagnetic vector meson decays into the lepton pair. The problem considered is well-known. We argue that it can be solved in the framework of $1/N_c$ expansion. An important step of our study is the gauge-invariant form of the Lagrangian describing the transition of the vector meson to a photon. Since in the processes under consideration the photon is far from its mass shell, this interaction does not vanish. One should recall that the standard VMD picture does not have the $V\to\gamma$ transitions through the meson loop.      
  
In our estimates, we have not calculated the self-energy diagrams of vector mesons. It is implied that the renormalization of vector-meson masses are already performed to their physical values, as it has been also assumed in \cite{Weise:96}. 
  
Our result is based on the effective meson Lagrangian of the NJL model and the $SU(3)\to SU(2)$ symmetry breaking mechanism which is implemented in the model, as like as in QCD, i.e., through the corresponding mass term of the free quark Lagrangian. The gap equation, which relates the masses of the current and constituent quark fields, makes it possible to trace the consequences of such a violation up to the vertices of the effective meson Lagrangian. We stress the importance of this step for the problem considered here.   

To summarize: we have demonstrated that in the framework of the $1/N_c$ approximation, the $\rho\to\pi\pi$ decay constant $g_\rho =6$. It is this value that should be also used in the theoretical description of two-lepton decays of neutral vector mesons. This finding supports the idea of vector meson universality, showing the importance of taking into account the contributions of the $1/N_c$ suppressed meson one-loop diagrams. These contributions do not exceed 30\% in comparison with the tree result for the $V\to l^+l^-$ decays and thereby bring the results of theoretical calculations into full agreement with experimental data and universality.

\section*{Acknowledgments}
The authors are grateful to A.\,B. Arbuzov for his interest in this work and useful discussions. The work of A.\,A. Osipov is supported by Grant from Funda\c{c}\~ ao para a Ci\^ encia e Tecnologia (FCT) through the Grant No. CERN/FIS-COM/0035/2019, and the European Cooperation in Science and Technology organisation through the COST Action CA16201 program. The work of K. Nurlan is supported by the JINR grant for young scientists and specialists No. 21-302-04, and the fund from the Science Committee of the Ministry of Education and Science of the Republic of Kazakhstan, Grant No. AP09057862.

\end{document}